\documentstyle[12pt]{article}
\begin{document}
\title{\bf Non-exponential decays and nonextensivity}
\author{G.Wilk$^{1}$\thanks{e-mail: wilk@fuw.edu.pl} and Z.W\l
odarczyk$^{2}$ \thanks{e-mail: wlod@pu.kielce.pl}\\[2ex] 
$^1${\it The Andrzej So\l tan Institute for Nuclear Studies}\\
    {\it Ho\.za 69; 00-689 Warsaw, Poland}\\
$^2${\it Institute of Physics, \'Swi\c{e}tokrzyska Academy}\\
    {\it  Konopnickiej 15; 25-405 Kielce, Poland}}  
\date{\today}
\maketitle

\begin{abstract}
We demonstrate that non-exponential decays of unstable systems
can be understood as yet another example of nonextensivity
encountered in many physical systems and as such can be 
characterized by the nonextensivity parameter $q$.\\   

[3ex]

\end{abstract}

\newpage

It is known that for the unstable system, for which the probability
$\gamma$ to decay in a unit of time is constant, its survival time 
probability is given by simple exponential law 
\begin{equation}
P(t) = \exp( - \gamma\, t ) . \label{eq:explaw}
\end{equation}
This means that our system is not really aging but with passing time
shows its persistency to decay. On the other hand, it is also widely
known \cite{FGR} that spontaneous decays of quantum-mechanical
unstable systems cannot be described by the pure exponential law
(neither for short nor for long times), i.e., one observes that for
them $ P(t)\, >\, \exp( - \gamma t )$ and, in practice, $P(t)$
exhibits the power-like behaviour of the type \cite{FGR}\footnote{The
ability to observe such nonexponential decays \cite{GW} and interest
in quantum version of the Zeno paradox \cite{ZENO} resulted in the
increase of interest in the problem of unstable quantum systems.}
\begin{equation}
P(t)\, \propto\, t^{- \delta} .\label{eq:Power}
\end{equation}
We would like to concentrate in this letter on the observation that
there is enormous variety of different physical phenomena which do
exhibit a power-like behaviour instead of naively expected
exponential one. They can be described in a most economical and
adequate way in terms of Tsallis nonextensive statistics \cite{T} 
by introducing only one new parameter $q$, known as the nonextensivity
parameter. The common feature of these phenomena is that they are
characterized either by long-range interactions and/or long-range
microscopic memories or by space-time (and/or phase-space)
(multi)fractal structure of the underlying physical process \cite{T},
or by intrinsic fluctuations of some parameters present in the system
under consideration \cite{WW}\footnote{This last point, which we
shall explore here, is receiving growing attention recently
\cite{BECK}.}.  In this approach one parameter power-like 
formula replaces the usual exponential (Boltzmann-Gibbs)
one\footnote{We would like to stress at this point the fact that 
eq. (\ref{eq:LBG}) is the exact solution of a simple nonlinear
ordinary differential equation of the type: $\frac{dy}{dt} = \frac{-
y^q}{\tau}$ with $y(t) = \frac{L(t)}{L(0)}$, as discussed in detail
in cf. \cite{Ta}.}:  
\begin{equation}
L_q(t)\, =\, \frac{2 - q}{\tau} \left[\, 1\, -
                   \, (1\, -\, q)\, \frac{t}{\tau}\,
                   \right]^{\frac{1}{1 - q}}\quad \stackrel{q
\rightarrow 1}{\Longrightarrow}\quad L_{q=1}(t)\, =\,
\frac{1}{\tau}\cdot \exp\left[\, -\, \frac{t}{\tau}\,\right] .
\label{eq:LBG} 
\end{equation}
The question we would like to address is whether non-exponentiality
of decays mentioned above can be expressed in the same approach, by
deviation of the suitably defined parametr $q$ from unity.\\

In \cite{WW} we have demonstrated that parameter $q$ (and power-like
distribution it characterizes) occurs whenever parameter $1/\tau$
of the original exponential distribution fluctuates according to
gamma distribution. In this case $q$ is given by the relative
variance of such fluctuations. We shall demonstrate now, using random
matrix approach \cite{M} that, indeed, the non-exponential decays can
emerge in a natural way from the possible fluctuations of parameter
$\gamma = 1/\tau$ in exponential distribution (\ref{eq:explaw}). To
this end let us start with discussion of some consequences of quantum
chaos, which emerge from applying random matrix hamiltonians,
elements of which follow some random probability distribution
\cite{M}. Namely, let us consider situation where we have $N$
intercoupled states with couplings between them described by
hamiltonian given in terms of $N\times N$ hermitian (and
rotatiotionally and time reversal invariant) matrix (with real matrix
elements): $H\, =\, \left[\, H_{ij}\, \right]_{N\times N}$, $(H_{ij}
= H_{ji})$. Suppose that one is interested in ensemble of such random
matrices with $P(H)$ being a probability of a given matrix $H$.
There is a number of propositions how to define $P(H)$ correspoding
to different models of randomness of matrix elements used \cite{R}.
However, in the limit of large matrices (large $N$) which are hermitian
and real, all of them lead to results analogous to those obtained by
assuming gaussian ensemble\footnote{The gaussian ensembles were first
discussed by E.P.Wigner \cite{G}. For analysis of the gaussian
ensemble considered here see \cite{G1}.}. This can be seen as
follows. Let $P(H)$ be such that: $(i)$ - elements $H_{ij}$ are
uncorrelated,  i.e.,  
\begin{equation}
P(H)\, =\, \prod^N_{i=1} \prod^N_{j\ge 1}\, P(H_{ij}), \label{eq:PH}
\end{equation}
$(ii)$ - distribution (\ref{eq:PH}) is invariant in respect to
arbitrary unitary transformation of the basic states (because initial
basic states $|i>$ used to define $H_{ij}$ have been chosen in
arbitrary way). Let us now look for distribution $P(C_i)$ of the
amplitudes $C_i$ for eigenvectors of $H$ assuming that our hermitean
ensemble is invariant in respect to the orthogonal change of the
basic vectors. Such ensemble is distributed in the probability space
with some density $P(C_1,\dots ,C_N)$ with $C^2_1 + C_2^2 + C_3^2 +
\cdots + C^2_N\, =\, 1$ and joint distribution function for $N$
amplitudes has the form
\begin{equation}
P(C_1,\dots ,C_N)\, =\, \frac{2}{\Omega_N}\, \delta\left( 1\, -\,
\sum^N_{i=1}C^2_i\right) \label{eq:JOINT}
\end{equation}
(where normalization constant $\Omega_N$ is equal to the volume of
$N$-dimensional sphere of unit radius and factor $2$ originates from
the fact that both positive and negative $C_i$ enter $P(C_1,\dots
,C_N)$). Integrating (\ref{eq:JOINT}) over all but one coordinates
$C_i$ gives the following distribution $P(C_i)$ of amplitude $C_i$ 
\begin{equation}
P(C_i)\, =\, \frac{\Omega_{N-1}}{\Omega_N}\, \left(1\, -\,
C^2_i\right)^{\frac{N-3}{2}}\, =\,
\frac{\Gamma\left(\frac{N+2}{2}\right)}{\sqrt{\pi}\Gamma
\left(\frac{N+1}{2}\right)}\, 
\left(1\, -\, C_i^2\right)^{\frac{N-3}{2}} , \label{eq:PCI2}
\end{equation}
which in the limit of large $N$ has the following form \footnote{By
using Stirling formula for ($N>>1$) leading to $\ln P(C_i) \simeq 
\frac{1}{2}\ln\frac{N}{2\pi} + \frac{N}{2}\ln\left(1-C^2_i\right)$,
and noticing that for large $N$ we have $C_i^2 << 1$.}
\begin{equation}
P(C_i)\, =\, \sqrt{\frac{N}{2\pi}}\, \exp\left(-
\frac{N}{2}C_i^2\right)\, =\, 
 \frac{1}{\sqrt{2\pi \left\langle C_i^2\right\rangle } }\,
 \exp\left( - \frac{C_i^2}{2 \left\langle C^2_i\right\rangle}\right) ,
 \label{eq:FINAL}
\end{equation}
where
\begin{equation}
\left \langle C_i^2\right \rangle \, =\, \frac{1}{N} 
\qquad {\rm and}\qquad 
\frac{\left\langle \left(C_i^2\right)^2\right\rangle\, -\, 
\left\langle C_i^2\right\rangle^2}{\left\langle
C_i^2\right\rangle^2}\, =\, 2 . \label{eq:VAR}
\end{equation}
However, for our purpose we are really interested in decay width
$\gamma_i$, which is proportional to the probability $C_i^2$ of
finding single component $|i>$ in the composite state. Its
distribution is given by
\begin{equation}
P(\gamma_i)d\gamma_i\, =\, 2\, P(C_i)\, |dC_i| , \label{eq:GAM}
\end{equation}
where factor $2$ reflects the fact that both positive and negative
values of $C_i$ contribute to the same value of $\gamma_i$. Because
$|dC_i|=\left| \frac{dC_i}{d\gamma_i}\right|d\gamma_i$ one gets
following distribution of such widths:
\begin{equation}
P(\gamma_i)d\gamma_i\, =\,
2P(C_i)\left|\frac{d\gamma_i}{d\gamma_i/dC_i}\right|\, =\,
\frac{1}{\sqrt{2\pi\gamma_i<\gamma_i>}} \, \exp\left( -
\frac{\gamma_i}{2<\gamma_i>}\right)\, d\gamma_i  \label{eq:PG}
\end{equation}
characteristic function of which is $\phi(z)\, =\, (1\, -\, i\,
2<\gamma_i>\, z )^{-1/2}$. In the case where contribution to the
width $\gamma$ comes from many channels, distribution we are
looking for is obtained by summing all contributions from the
particular channels of widths $\gamma_i$. For $\nu$ different
independent channels with the same mean values $\gamma_0 =
<\gamma_i>$ the corresponding characteristic function will be
$\phi_{\nu}(z)\, =\, (1\, -\, i\, 2\gamma_0\, z)^{-\nu/2} $ (with
$<\gamma> = \nu \gamma_0$), i.e., the corresponding distribution
function will be gamma function of the form 
\begin{equation}
P_{\nu}(\gamma) d\gamma\, =\, \frac{1}{\Gamma\left(\frac{\nu}{2}\right)}\,
                           \left(\frac{\nu}{2<\gamma>}\right)\, 
           \left(\frac{\nu \gamma}{2<\gamma>}\right)^{\frac{\nu}{2}-1}\, 
           \exp\left(- \frac{\nu \gamma}{2<\gamma>}\right) . \label{eq:FINALG}
\end{equation}
This completes our proof that, as a results of composition of
contributions from the $\nu$ channels, we obtain distribution of 
the widths of the decays $P_{\nu}(\gamma)$ in the form of gamma
distribution, fluctuations of which are characterized by the relative
variance of the form
\begin{equation}
\frac{\left\langle (\gamma \, -\,
<\gamma>)^2\right\rangle}{<\gamma>^2}\, = \, \frac{2}{\nu} 
\end{equation}
and they decrease with growing $\nu$ \footnote{The (\ref{eq:FINALG})
is for the integer values of $\nu$ encoutered here identical to the
$\chi^2$ distribution obtained in \cite{BECK}.}.\\

We have therefore demonstrated that, indeed (at least in the above
example), the fluctuations of the life times $\tau = 1/\gamma$ around
mean value $\tau_0 = 1/<\gamma>$ are described by gamma distribution
of the form (\ref{eq:FINALG}) of the variable $1/\tau$. According to
our philosophy such fluctuations of parameter $\gamma$ in the
exponential distribution (\ref{eq:explaw}) lead therefore directly to
the L\'evy distribution of the form \cite{WW,BECK} 
\begin{equation}
L_q(t,\tau_0)\, =\, \int_0^{\infty} d\left(\frac{t}{\tau}\right)\,
P_{\nu}\left(\frac{t}{\tau}\right)\, \exp\left(
-\frac{t}{\tau}\right)\, =\, \frac{2-q}{\tau_0}\, \left[1\, -\,
(1-q)\frac{t}{\tau_0}\right]^{\frac{1}{1-q}} \label{eq:Pq}
\end{equation}
with nonextensivity parameter $q$ equal to $q = 1 + \frac{2}{\nu}$.\\

We shall close with some remarks on chaotic systems and mixing of
configurations. Notice first that distribution
$P_{\nu}(\gamma)d\gamma$ is known for $\nu =1$ in the nuclear physics
as the so called Porter-Thomas distribution \cite{PT}
\footnote{Deviations from this distribution, which can be attributed
to small $N$ involved, have been discussed recently in \cite{PT1}.}
and describes in a satisfactory way the widths of resonances observed
experimentally \cite{EXP}. On the other hand the many channel case of
$\nu > 2$ corresponds, for example, to the case of total widths for
gamma emision or for overbarrier fission. Model using
random matrices corresponds to the extreme case of total mixing
between allowed degrees of freedom. Such mixing of configurations is
realized in the case of compound nucleus (where energy of
excitation is divided among many degrees of freedom). It means that
component representing one-particle motion in reference to the whole
nucleus is decomposed into wave function of a large number of
different resonance states. In the Fermi gas model the compound
nucleus corresponds to a particularly strong mixing of different
configurations, both corresponding to the stationary and to the
resonance states. Such emergence of the mixing of configurations,
which is leading to the compound nucleus, can be understood as 
effect of an increase of importance (paralel to the increase of the
excitation energy of the system) of small perturbations in the
one-particle motion. It happens because in this case the number of
exciting particles increases (as increases the number of possible
interactions) and excitation energy per single particle
increases as well (the density of final states accessible in the
collisions increases) \footnote{In this respect there is a whole list
of problems which could possibly be considered, for example, the
problem of mixed states in the quantum physics, connection with the
composite biochemical systems for which fluorescence time
distributions are known to be non-exponential (and are usually 
described by a composition of a number of exponents).}.\\

One should stress also that random matrix approach is applied in the
chaotic quantum systems \cite{Q} and it seems to be the
characteristic feature of such quantum systems, which in the
classical limit show chaotic behaviour. It is connected with the fact
that in such systems there is no symmetry, i.e., there is no
degeneracy and no selection rules which would exclude interactions of
some particular states \footnote{For application of random matrices
to description of chaos in quantum systems see \cite{APPL}}.\\

We are grateful for fruitful discussions with Prof. C. Tsallis. The partial
support of Polish Committee for Scientific Research (grants 2P03B 011
18 and  621/E-78/SPUB/CERN/P-03/DZ4/99) is acknowledged.

\end{document}